\documentclass[12pt,preprint,usenatbib]{aastex}
\def\mpc{h^{-1} {\rm{Mpc}}}

\def\kms {\rm{km~s^{-1}}}
\def\ub {^{0.1}u}
\def\gb {^{0.1}g}
\def\rb {^{0.1}r}
\def\ib {^{0.1}i}
\def\zb {^{0.1}z}
\shorttitle{Luminosity function of galaxies in groups}
\shortauthors{Zandivarez, Mart\'\i nez & Merch\'an}
\begin{document}
\title{On the luminosity function of galaxies in groups in the Sloan Digital Sky Survey}
\author{Ariel Zandivarez, H\'ector J. Mart\'\i nez and Manuel E. Merch\'an}
\affil{ Grupo de Investigaciones en Astronom\'{\i}a Te\'orica y Experimental (IATE)\\
Observatorio Astron\'omico, Universidad Nacional de C\'ordoba,\\
Laprida 854, X5000BGR, C\'ordoba, Argentina}
\affil{Consejo Nacional de Investigaciones Cient\'{\i}ficas y T\'ecnicas (CONICET),\\
Avenida Rivadavia 1917, C1033AAJ, Buenos Aires, Argentina}
\email{arielz@mail.oac.uncor.edu, julian@mail.oac.uncor.edu, manuel@mail.oac.uncor.edu}

\begin{abstract}
Using galaxy groups identified in the Fourth Data Release of the Sloan Digital 
Sky Survey (SDSS), we compute the luminosity function for several subsamples of 
galaxies in groups.  In all cases, the luminosity functions are well described 
by Schechter functions, down to the faintest magnitudes we probe, 
$M_{\rb}-5\log(h)\sim-16$.
For the general luminosity function of
galaxies in groups in the five SDSS bands, we observe that the characteristic magnitude 
is brighter in $\sim 0.5$ magnitudes compared to those obtained for field galaxies by 
\citet{blantonlf}. 
Even when the observed faint end slope is steeper in galaxy groups, 
it is statistically comparable with the field value. 
We analyze the dependence of the galaxy luminosity function with system masses 
finding two clear trends: a continuous
brightening of the characteristic magnitude and a steepening of the faint end slope as
mass increases. The results in $\gb$, $\rb$, $\ib$ and $\zb$ bands show the same behavior.
Using the $u-r$ color to split
the galaxy sample into red and blue galaxies, we show that the changes observed 
as a function of the system mass are mainly seen in the red, passively evolving, galaxy 
population, while the luminosities of blue galaxies remain almost unchanged with mass.    
Finally, we observe that groups having an important luminosity difference 
between the two brightest galaxies of a system show a steeper faint end slope 
than the other groups. Our results can be interpreted in terms of galaxy mergers
as the main driving force behind galaxy evolution in groups. 
\end{abstract}


\section{Introduction} 
To understand the processes that govern galaxy formation and evolution,
detailed information must be collected about the behavior of the different
galaxy populations under different conditions. A common and useful way to
achieve this, is the study of galaxy luminosities and their variation
with the environment. The most suitable statistical tool to perform this kind 
of analysis is the luminosity function (LF) of galaxies. 
This function describes the distribution of luminosities of a given population of 
galaxies and, in most cases, it can be parametrized
by a function with two parameters excluding the normalization (\citealt{schechter76}), 
the characteristic absolute magnitude $M^{\ast}$ and the faint end slope $\alpha$. 
The results obtained from the analysis of these parameters are among the most 
interesting issues in extragalactic astronomy.

Prior to 2000 the LF has been computed for galaxies in 
the field, groups and in clusters of galaxies (see for instance 
\citealt{mhg94,lin96,zucca97,lopcruz97,valotto97,ratclif98,mvl98,rauzy98,tren98}). 
Two lines of thought have arisen from these 
works: one of them states that the LF depends on the environment while the other 
supports the idea of an universal LF. 
The dependence of the LF with the environment was
proposed by some authors due to the very steep faint end slope found for the LF in 
clusters of galaxies which was interpreted as an excess of dwarf galaxies 
relative to the field.
Nevertheless, in order to have more reliable results on this matter,
large samples of galaxies with high quality photometric and 
spectroscopic information were needed.

Since the advent of the large surveys of galaxies, such as the Sloan Digital 
Sky Survey \citep{sdss} and the Two degree Field Galaxy Redshift Survey (2dFGRS, 
\citealt{2df}), much better determinations of the LF have been obtained 
\citep{blanton01,blantonlf,blanton05,norb02,madgwick,trentu02,m02b,cz03,eke04b}.
Most of these works agree with a flat LF of galaxies in the field ($\alpha \sim -1$) 
meanwhile a brighter characteristic magnitude $M^{\ast}$ and a steeper faint end slope 
$\alpha$ have been found in galaxy systems.
Analyzing  a sample of rich clusters of galaxies obtained from a cross-correlation
between the Sloan Digital Sky Survey and the Rosat All Sky Survey, \citet{pop05} have
found that the very faint end slope ($M_r-5\log(h)>-17$) is remarkably steep 
($\alpha \sim -2$) on these environments.  
Similar results have been found by \citet{gonz05} when lower density environments
as galaxy groups are analyzed. It should be noted that both works have been carried out 
using statistical background subtraction methods owing to the lack of spectroscopic 
information for faint galaxies. These methods are sensitive to the background
computation \citep[see their Table 2]{pop05} and to the presence of structures 
along the line of sight \citep{val}, therefore, they are less reliable than those based 
only upon spectroscopic information.

Groups of galaxies are very interesting objects in the universe, since
they are the most common systems of galaxies and the evolution of galaxies inside
them plays an important role at early stages of cluster galaxies evolution.
How important is the variation of galaxy luminosities with the main physical
properties of these systems? How does the environment affect the different galaxy 
populations?. Large statistical samples of galaxy groups are needed in order 
to these questions be addressed.

The largest galaxy redshift survey at the present is the Fourth Data Release of the
Sloan Digital Sky Survey (hereafter DR4; \citealt{dr4}). 
This catalog covers a very wide area on the sky, 
has high quality information in five broadbands and comprises two times 
the number of galaxies in the current Final Release of the 2dFGRS.
The huge size of the Sloan galaxy catalog will provide us of the largest
galaxy group catalog at the present with very important photometric and
spectroscopic information.  

The main purpose of this work is to compute the LFs of galaxies in groups
under different conditions in order to understand the behavior of 
galaxies in intermediate density environments and consequently 
provide important clues on galaxy formation and evolution.

The layout of this paper is as follows. In section 2 we describe the
galaxy sample and the group identification. The full analysis of 
LFs as a function of the band, mass ranges, color
and the brightest galaxy is in section 3. Finally in section 4 we 
summarize and discuss the results. 

\section{Sample}
The Sloan Digital Sky Survey (SDSS) has validated and made publicly available its
Fourth Data Release (DR4; \citealt{dr4}) which is a photometric and spectroscopic 
survey constructed with a dedicated $2.5 \ {\rm m}$ telescope at Apache 
Point Observatory in New Mexico. 
The SDSS DR4 consists of $6670~ {\rm deg}^2$ of five-band, $u \ g \ r \ i \ z$, 
imaging data and 673,280 ($ 4783~ {\rm deg}^2$) spectra of galaxies, quasars and stars. 
The DR4 Main Galaxy Sample (MGS; \citealt{mgs}) includes roughly  
400,000 galaxies with redshift measurements up to $z\sim0.3$  and an upper apparent 
magnitude limit of 17.77 in the $r$ band.

Galaxy groups used in this work have been identified in the MGS of DR4 
using the same procedure as in \citet{mz05}. 
The method consists in using a friend-of-friend algorithm similar to
that developed by \citet{hg82}. The algorithm links galaxies ($i,j$) which satisfy that 
$D_{ij}\le D_0 \ R(z)$ and $V_{ij}\le V_0 \ R(z)$ where $D_{ij}$ is the projected distance and
$V_{ij}$ is the line-of-sight velocity difference. The scaling factor $R(z)$ is introduced
in order to take into account the decrement of the galaxy number density due to the
apparent magnitude limit cutoff (see eq. 5 in \citealt{hg82}). 
We have adopted a transversal linking length $D_0$ 
corresponding to an overdensity of $\delta \rho/\rho=80$, 
a line-of-sight linking length of $V_0=200~\kms$ and a fiducial distance 
of $10~\mpc$.
As in \citet{mz05}, we have also carried out an improvement of the 
rich group identification.  This improvement consists in performing a second 
identification on galaxy groups with at least ten members in order to split 
merged systems or to eliminate spurious member detections.
This second identification uses a higher density contrast, $\delta \rho/\rho \sim 315$,
which produces a more reliable group identification (see \citealt{diaz05}). 
Group center locations for these groups have been improved using an iterative 
procedure developed by \citet{diaz05}.
The procedure defines a new group center estimation by using the projected 
local number density of each member galaxy for weighting their group center 
distances, and then iterates this estimation by removing galaxies beyond a 
given distance. 
The iteration follows until the center location remains unchanged. Since the method 
needs to compute the projected local number density with five galaxies, this procedure
can only be applied to group with at least ten members.  

Finally, the group sample comprises 14004 groups with at least 4 members,
adding up to 85728 galaxies. Mean properties of groups in this sample
are similar to those of the DR3 group catalog as quoted by \citet{mz05}.
We obtain a median velocity dispersion, virial mass and radius of
$232 \ \kms$, $3.9\times10^{13} \ h^{-1} \ {\cal M}_{\odot}$, and $1.11 \ \mpc$ respectively.

The magnitudes we use in this work are the Petrosian ones, and
have been corrected for Galactic extinction following \citet{sch98}.
Absolute magnitudes have been computed assuming $\Omega_0=0.3$, 
$\Omega_{\Lambda}=0.7$ and a Hubble constant $H_0=100~ h~ {\rm km~s^{-1}~Mpc^{-1}}$,
and $K-$corrected using the method of 
\citet{blantonk}\footnote{{\tt kcorrect} version 4.1}. 
We have adopted a band shift to a redshift $0.1$, 
i.e. to approximately the mean redshift of the Main Galaxy Sample of SDSS, as suggested
by \citet{blantonk}. We have also included evolution corrections to this
redshift for each galaxy of the form $({\rm z}-0.1)Q$ where $Q$ varies with the band: 4.22, 
2.04, 1.62, 1.61 and 0.76 for $u$, $g$, $r$, $i$, $z$ bands respectively \citep{blantonlf}.
Throughout this work we will refer to these shifted bands 
as $^{0.1}u \ ^{0.1}g \ ^{0.1}r \ ^{0.1}i \ ^{0.1}z$. All magnitudes are in the AB system.

\section{Luminosity function of galaxies in groups}
Following the comparative study of different luminosity function estimators 
by \citet{w97}, we have chosen two methods for
computing luminosity functions: the non-parametric $C^-$
method \citep{lb71,cho87} and the STY method \citep{sty79} for fitting Schechter
function parameters. Among the non-parametric methods, the $C^-$ is the most robust 
estimator, being less affected by different values of the faint end slope
of the Schechter parametrization and sample size. Meanwhile, the STY 
is reliable for fitting analytic expressions without binning the data.

Since we are only interested in studying the LF shape for different
subsamples of galaxies in groups, all LFs in this paper are shown in 
arbitrary units.

\subsection{The luminosity function in the SDSS bands}

For computing luminosity functions, we have adopted the same apparent 
magnitude cut-offs as in \citet{blantonlf}.
Given that $K-$corrections in each band are reliable only in a given redshift range,
we have also introduced redshift cut-offs depending on the band  
(Table \ref{sty}, see \citealt{blantonk}). 

\begin{figure}
\plotone{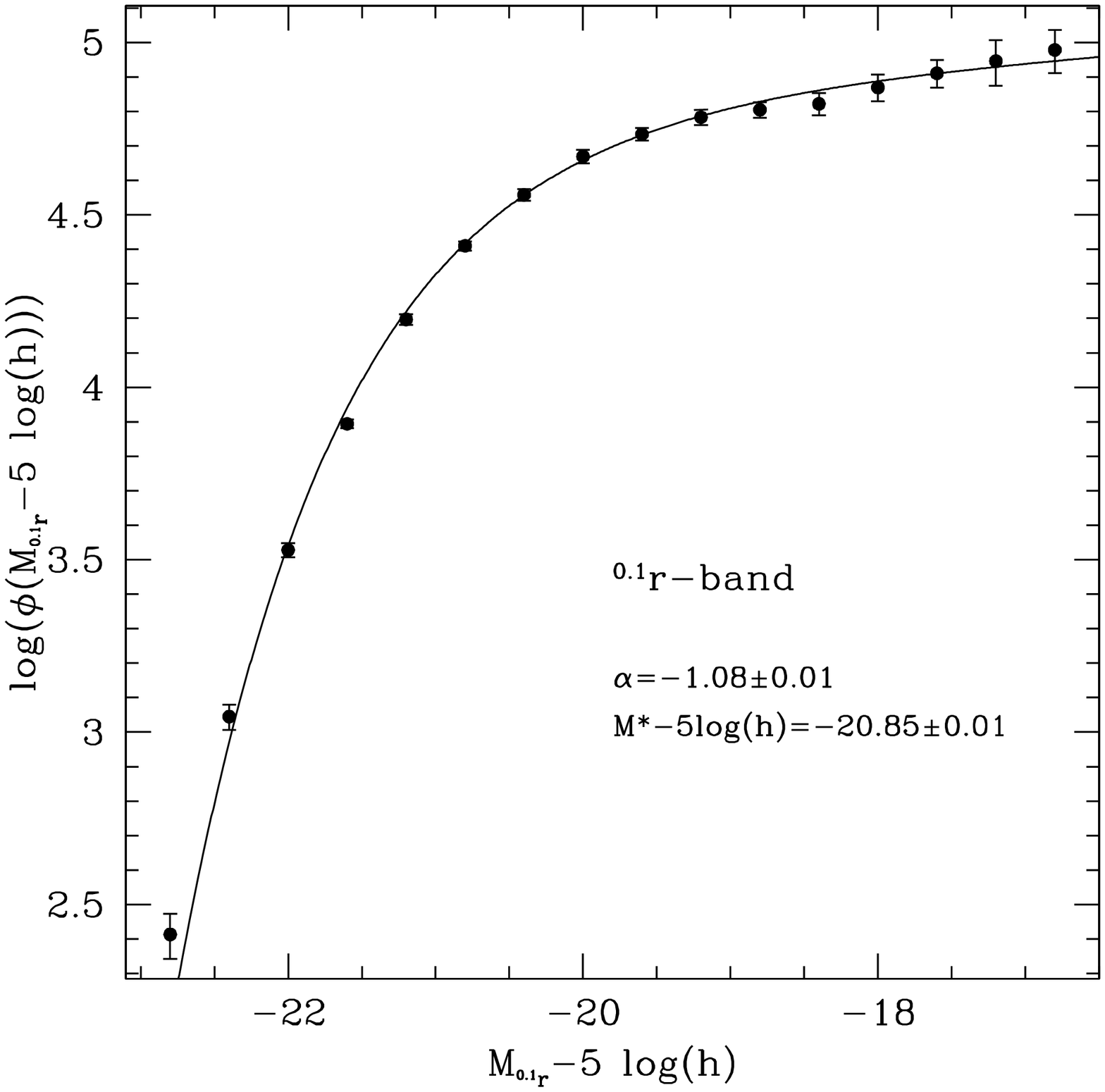}
\caption{
The $^{0.1}r$ band luminosity function of galaxies in groups in 
arbitrary units. Solid line shows the best-fitting Schechter 
function (see labels). Error bars were computed using the 
bootstrap resampling technique.
}
\label{lf}
\end{figure}

\begin{figure}
\plotone{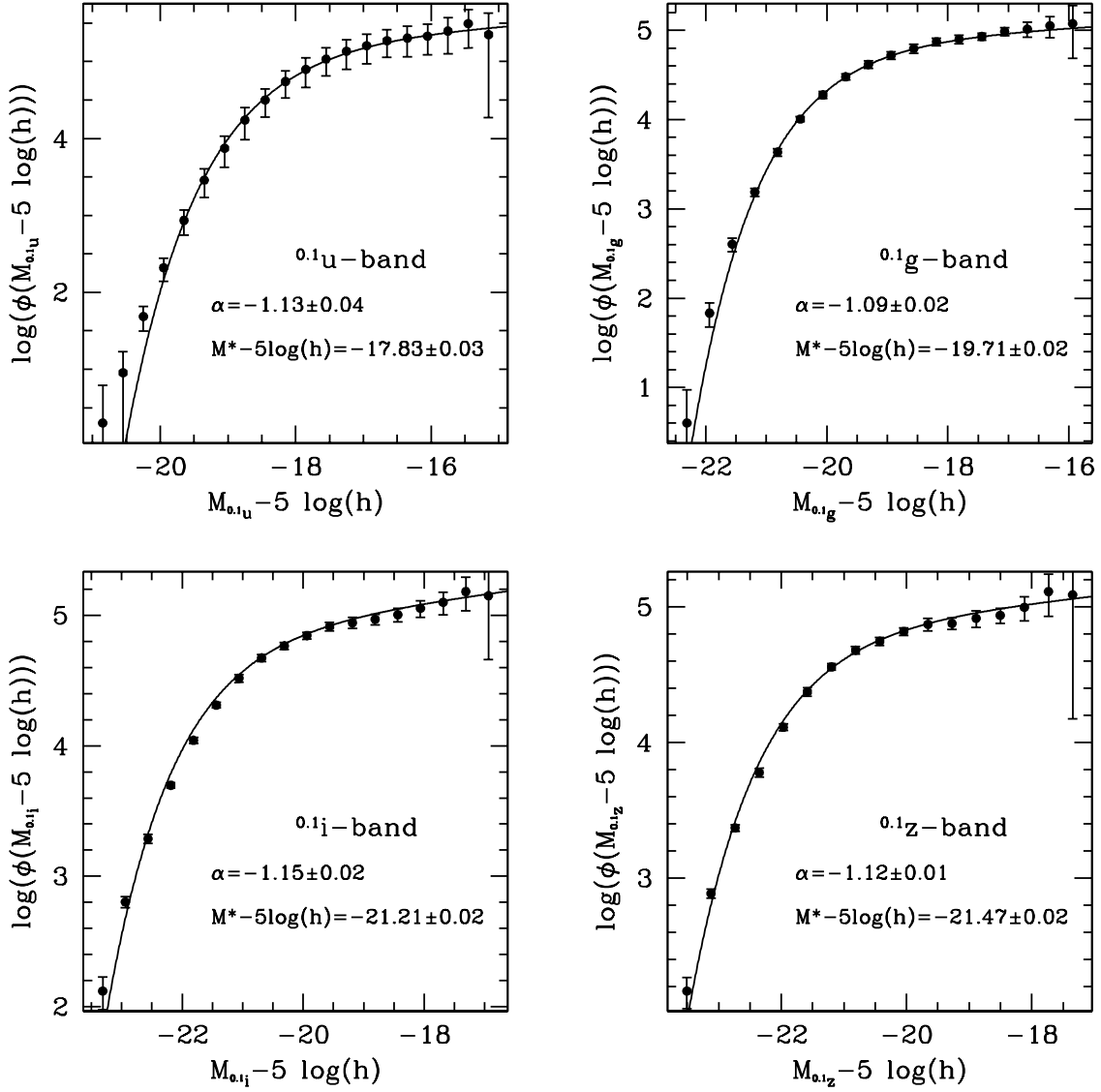}
\caption{
Luminosity functions of galaxies in groups in different bands
(arbitrary units).
Solid lines in each panel show the best-fitting Schechter 
functions (see labels). Error bars were computed using the 
bootstrap resampling technique.
}
\label{lfesplit}
\end{figure}

The $\rb$ band LF for galaxies in groups is shown in Figure \ref{lf},
along with the best Schechter function fit. The LF for the other bands
and the corresponding fits are displayed in Figure \ref{lfesplit}
\notetoeditor{Figures 2, 3 and 6 should span 2 columns}.
Best fit Schechter parameters can be found in Table \ref{sty}.
It is noticeable from these figures that Schechter functions provide a good
description of the LF for all bands. 

When comparing our results with those of \citet{blantonlf} (their Table 2) 
for the LFs of all galaxies in DR1's MGS, we observe that, 
with the exception of the $\ub$ band, the characteristic magnitudes, 
$M^{\ast}-5\log(h)$ are brighter for galaxies in groups. 
These observed brightening of $M^{\ast}$ ranges from $0.29$ in the $\zb$ band to 
$\sim0.5$ magnitudes in the $\gb$, $\rb$ and $\ib$, with more than $7\sigma$ 
significance in all cases. 
At the same time, group galaxies show a small steepening of the faint end slope, 
$\alpha$.  The steepening is more pronounced for bluer bands, $\ub,~\gb$, where
a change of $\sim0.2$ in $\alpha$ is observed. Nevertheless, these differences
are not very significant, since all of them lie within $\sim 2\sigma$.
These results are in agreement with those obtained by \citet{m02b} for galaxies 
in groups identified in the 100K release of the 2dFGRS, in the $b_J$ band. 
They found that the characteristic magnitude is brighter for galaxies in groups
respect to the LF obtained for field galaxies by \citet{norb02},
while the faint end slope for galaxies in groups was statistically comparable with 
the field value. 

In hierarchical models for galaxy formation and evolution,
the frequency of mergers increase in intermediate mass systems
such as groups. Therefore, in this scenario, galaxies in groups and clusters
are expected to be typically brighter than in the field, 
resulting in a brighter $M^{\ast}$. This could explain the
differences between group and field galaxies.
On the other hand, the $\ub$ band result, is not unexpected
since this band is closely related to the current star formation, and 
galaxies in systems are likely to have lower or suppressed star formation
rates (see for instance \citealt{m02}). 

\subsection{The group mass dependence of the galaxy LF}

In this subsection we study how the LF depends on group masses.
A previous analysis on this matter was done by \citet{eke04b}
using the 2dFGRS Percolation-Inferred
Galaxy Group catalog (2PIGG; \citealt{eke04a}). They split their group sample
into three mass bins and compute the corresponding LFs using the standard 
$1/V_{\rm max}$ and the STY methods. Though their high mass bin is statistically
poor and does not follow the same trend as the lower mass ones, 
they conclude that $M^{\ast}$ and $\alpha$ decrease when mass increases.

Given the large number of groups we have identified in DR4, we expect
to obtain a more detailed description on this subject.  
We have split the full sample of $z\le0.22$ groups into 6 mass bins
defined to have roughly the same number of groups ($\sim 2200$).
The details of this selection can be seen in Table \ref{stymass}.
The large number of galaxies in the samples allows the determination
of LFs with a high level of confidence, even when we are splitting the 
sample into several bins.
  
\begin{figure}
\plotone{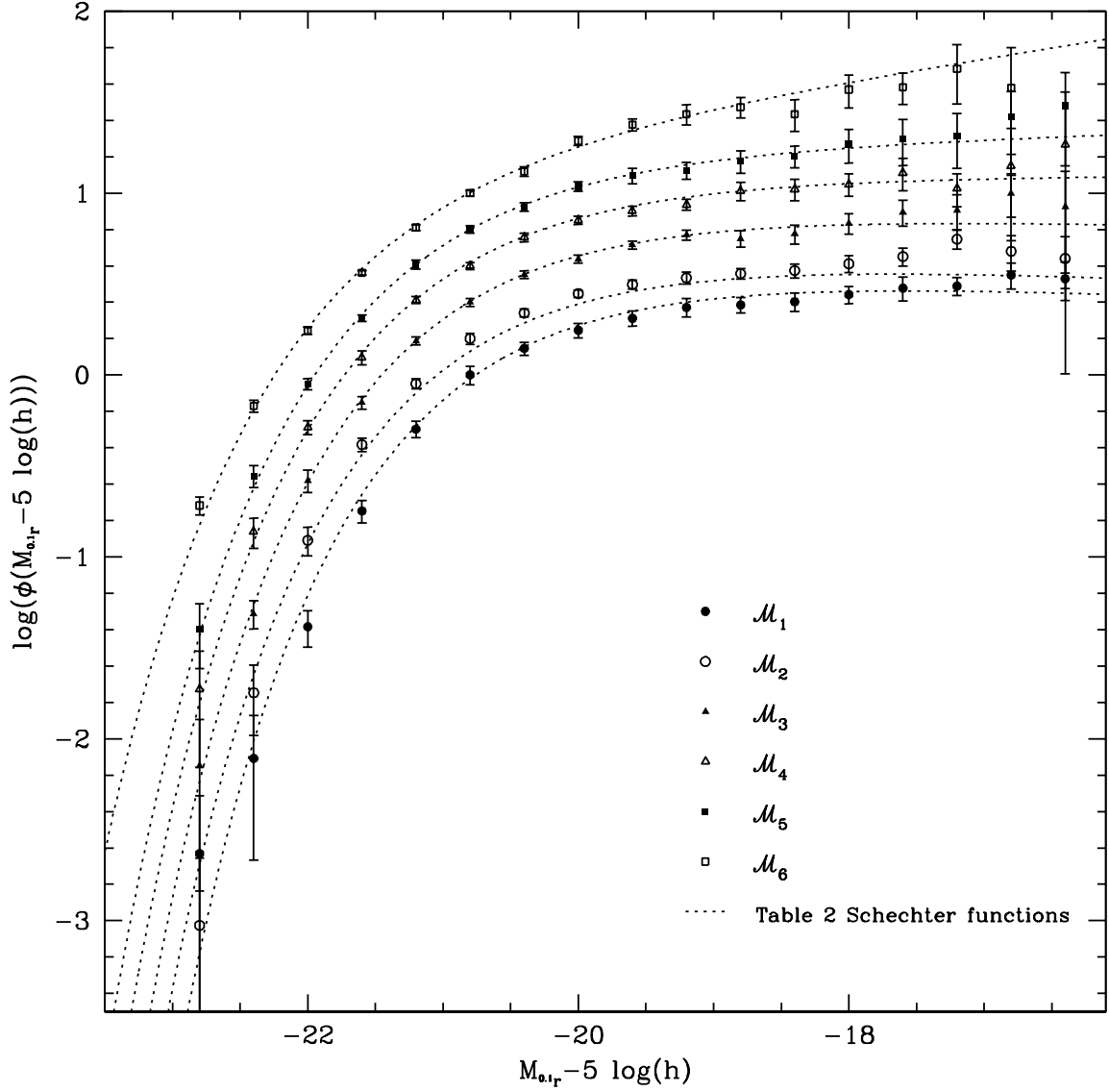}
\caption{
$^{0.1}r$ band luminosity functions of galaxies in groups 
in arbitrary units for different mass ranges (see Table \ref{stymass}). 
Dotted lines show the best-fitting Schechter 
functions. Error bars were computed using the
bootstrap resampling technique.
}
\label{lfmasa}
\end{figure}

\begin{figure}
\plotone{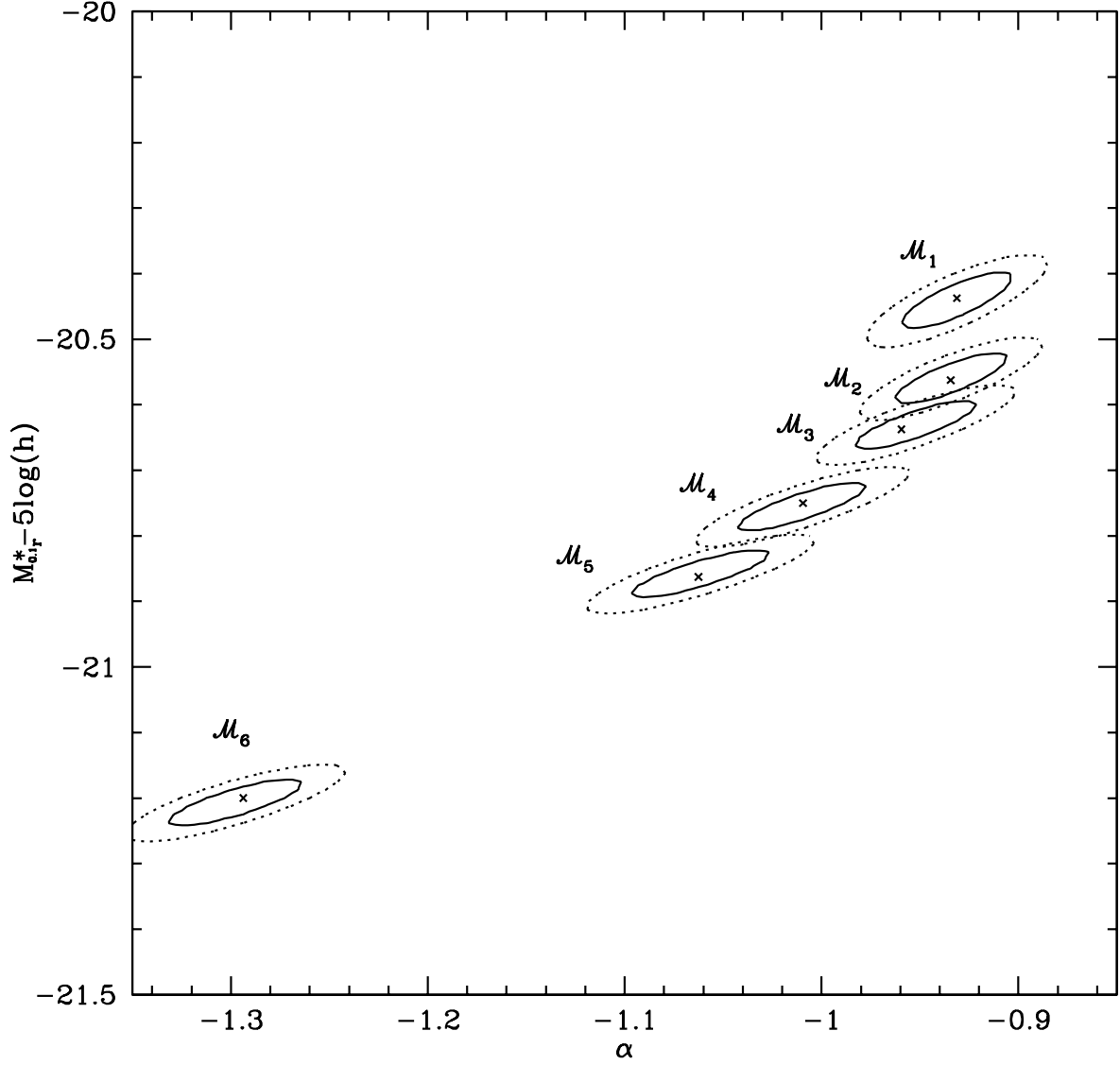}
\caption{
The best-fitting Schechter parameters of 
the $^{0.1}r$ band LF for different mass ranges. 
The $1\sigma$ and  $2\sigma$ confidence 
ellipses are also shown.
}
\label{isomasa}
\end{figure}

\begin{figure}
\plotone{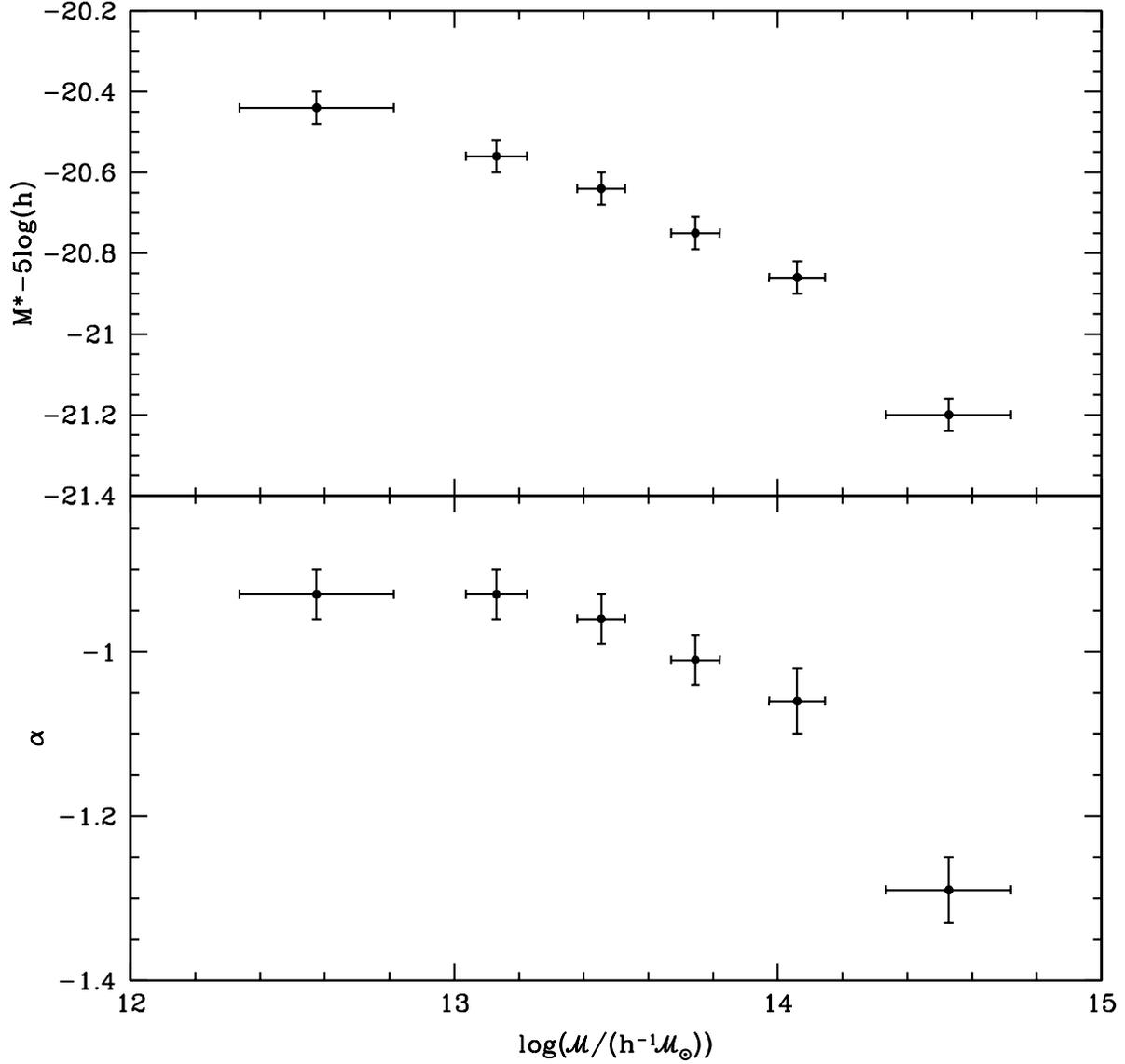}
\caption{
The best-fitting Schechter parameters $M^{\ast}-5\log(h)$ 
(upper panel) and $\alpha$ (lower panel) for
the $^{0.1}r$ band LF as a function of the median mass. 
The mass error bars are the semi-interquartile range, whereas
the errors in the Schechter parameters are the projections 
of 1$\sigma$ joint error ellipse onto each axis.
}
\label{amm}
\end{figure}

In Figure \ref{lfmasa} we show the resulting $\rb$ band LF for each mass bin.
As it can be seen from Figure \ref{lfmasa}, the LFs are well fitted
by Schechter functions for all masses (see parameters in Table \ref{stymass}).
In order to simplify the reader interpretation of the results, we 
show the values of $M^{\ast}$ {\it vs.} $\alpha$ (Figure \ref{isomasa}) 
and $M^{\ast}$ and $\alpha$ as a function of the mass (Figure \ref{amm}).  
Figure \ref{isomasa} displays the $1\sigma$ and  $2\sigma$ confidence 
ellipses according to STY computations. Y-axis error-bars in Figure \ref{amm}
are the projections of the 1$\sigma$ joint error ellipse onto each axis
of Figure \ref{isomasa}, while x-axis error-bars are the semi-interquartile
range of the median mass.
There are two clear and continuous trends: a brightening of $M^{\ast}$ and a 
simultaneous steepening of $\alpha$ as mass increases. $M^{\ast}$ changes in
$\sim 0.75$ magnitudes while $\alpha$ varies in 0.4, over two orders of 
magnitude in group mass.
We have done the same calculations for the other SDSS bands.
The best-fitting Schechter parameters are in Table \ref{stymassband}.
We found the same global trends in all bands, with some differences
for the $\ub$ band. In this case we observe a brightening of $M^{\ast}$
of $\sim 0.4$ magnitudes, significantly smaller than the $\sim0.75$ magnitudes 
value for the other bands.  
This is not unexpected, since the $\ub$ band is sensitive to star formation,
i.e., $\ub$ luminosity function is a poor indicator of the underlying mass 
distribution. The faint end slope variation with mass is roughly the same, $\sim0.5$,
for all bands.

A similar result was found by \citet{croton} analysing the variation of the
of the LFs of galaxies in the 2dFGRS with the environment. They study 
the dependence of the LFs with the density contrast estimated within $8 \mpc$ sphere
observing a smooth variation of the Schechter parameters in density
environments ranging from voids to clusters.
It should be taken into account that this parameter can not be directly 
related with group masses since the virial mass describes a closer 
environment than the corresponding to $8 \mpc$ density contrast, even so,
their results shows the same trends than the described in the previous
paragraph.

Our results means that, as system mass increases, the characteristic
luminosity of galaxies increases. 
Regarding the steepening of the faint end slope with mass, 
at least two possibilities arise: an important fraction of bright, 
$M_{\rb}\lesssim-18$, galaxies gets brighter, 
then the `knee' of the Schechter function becomes less pronounced and this
gives a steeper $\alpha$, or, there are some physical mechanisms that 
increase the number of faint, $M_{\rb}\gtrsim-18$, galaxies. 
Therefore, there exist some processes that enhance
galaxy brightness for bright galaxies and possibly some other processes that
increase the number of faint galaxies. The effect of these processes becomes more
noticeable for massive systems. 
Merging and galactic cannibalism are likely to be responsible of 
producing brighter galaxies. As the result of a tidal interaction between
two galaxies, the massive counterpart can get brighter while the less
massive diminishes its luminosity.
On the other hand, processes involving the interaction between
galaxies and the intra-system environment, such as ram pressure, 
are important for massive systems. It results in the loss of gas
in less bound galaxies, drastically reducing the star formation.

\subsection{The LF for the Red and Blue sequences in groups}

\begin{figure}
\plotone{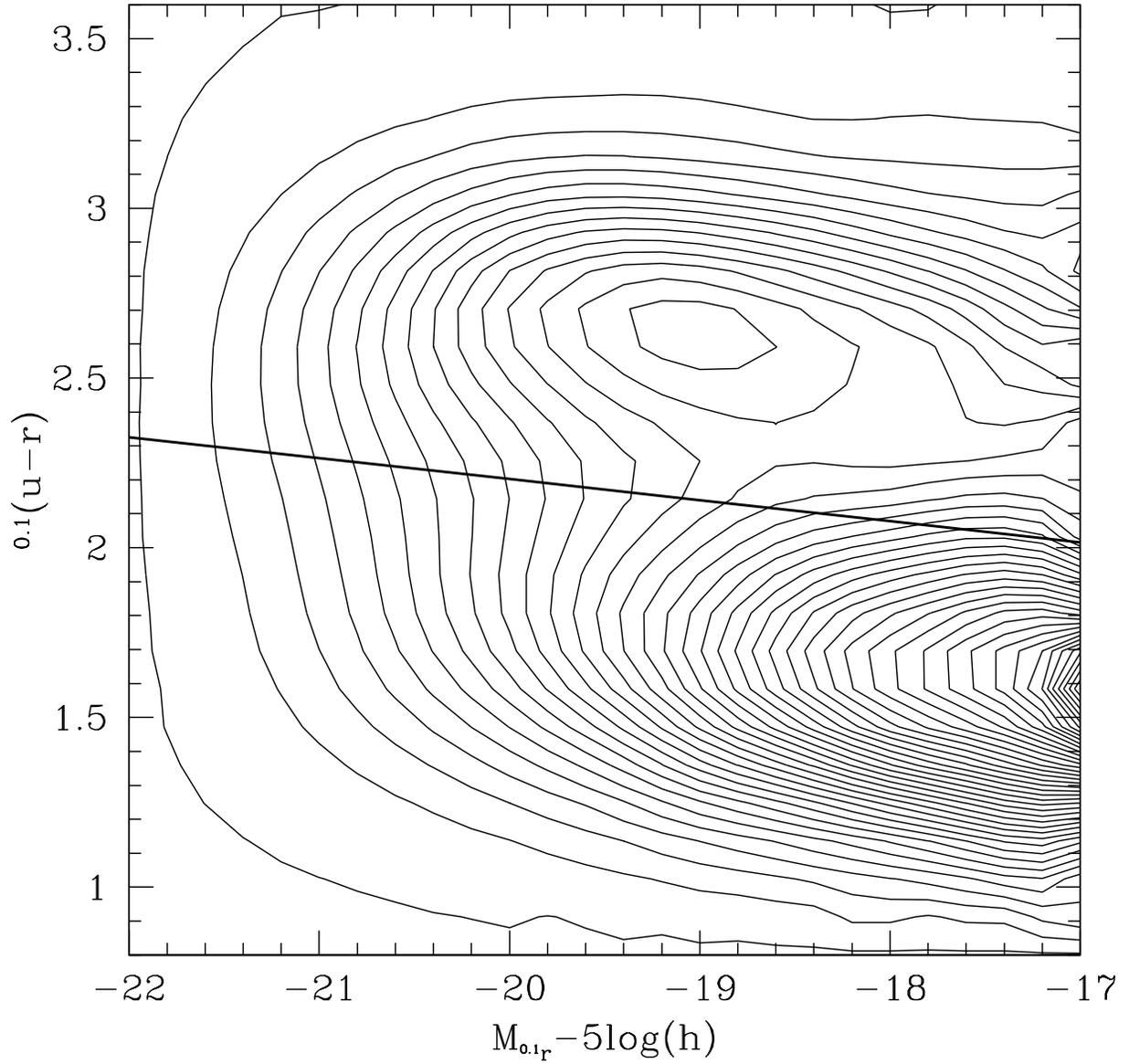}
\caption{
Color-magnitude distribution corrected for incompleteness. 
The contours are on a linear scale in number density.
The thick straight line is the divider point of the bimodal
distribution (see Equation \ref{cutoff}).
}
\label{hiscol}
\end{figure}

In a previous work, \citet{strat01} found that the color distribution of galaxies
can be approximated by a bimodal function, i.e. by the sum of two normal Gaussian 
functions. This behavior can be explained by two different formation processes which
generate two galaxy populations with different average colors.
The most common choice is to adopt the $u-r$ color to split the galaxy distribution
into two different populations. There are several works in the literature
that use this color distribution to distinguish between two galaxy populations. 
Recently, \citet{baldry03} have shown that it can not be chosen an unique 
color divider point since this point depends on absolute magnitude. 
So, in order to divide the sample of galaxies in groups into two different populations
we have parametrized the relation between the color divider point and the absolute 
magnitude in the $\rb$ band. To do so, we firstly use the MGS and split it into
absolute magnitude bins of width $0.5$. 
For each absolute magnitude bin, we fit to the $K+E-$corrected $^{0.1}(u-r)$ color 
distribution the sum of two Gaussian functions. The color divider point
is estimated as the intersection point between both Gaussian functions. 
Finally, we fit a straight line to the color divider points as a function of the
$\rb$ absolute magnitude. The resulting linear relation is
\begin{equation}
C_{\rm cut}(M_{\rb}) = -0.062 (M_{\rb}+18) + 2.078.
\label{cutoff}
\end{equation}
Figure \ref{hiscol} shows the $1/V_{\rm max}$ weighted color-magnitude distribution 
for the MGS. 
The straight line shows the color divider point as a function of absolute magnitude
according to Equation \ref{cutoff}.
Using this function we split the sample of galaxies in groups into two subsamples of
Red, $^{0.1}(u-r)>C_{\rm cut}(M_{\rb})$, and Blue, $^{0.1}(u-r)<C_{\rm cut}(M_{\rb})$, 
galaxies. 

\begin{figure}
\plotone{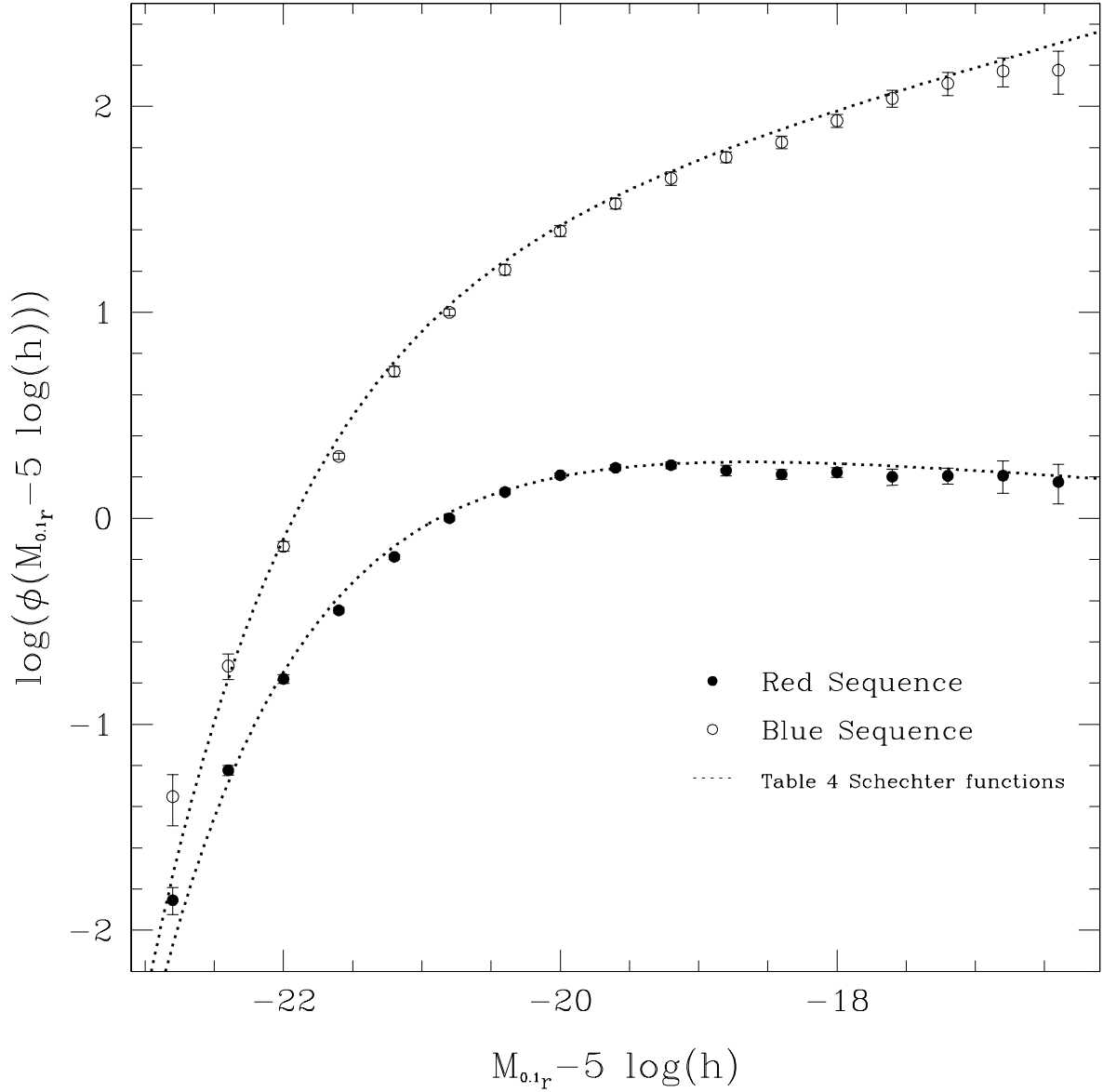}
\caption{
$^{0.1}r$ band luminosity function of the Red (upper panel)
and Blue (lower panel) sequences in groups in arbitrary units. Dotted lines show
the best-fitting Schechter functions. 
Error bars are computed using a bootstrapping technique.
}
\label{lfcolor}
\end{figure}

The $\rb$ band LFs for the Red and Blue sequences of galaxies in groups
are shown in Figure \ref{lfcolor} together with the respective Schechter 
best-fitting parameters (see also Table \ref{stymasscolor}). 
When comparing our results with field values by \citet{baldry03},
we observe that the Red sequence in groups has a brighter characteristic
magnitude and a slightly steeper faint end slope. Taking into account error-bars
and the band shift, the Red sequence in the field and in groups have comparable LFs.
On the other hand, there are some differences between the Blue sequence
in the field and in groups. We find a brighter $M^{\ast}$
and a steeper $\alpha$ in groups.
\citet{baldry03} find $M^{\ast}_r-5\log(h)=-19.82\pm0.08$ for the Blue sequence, 
which means a difference of $\sim 0.8$ magnitudes in $M^{\ast}$.
The Blue sequence faint end slope is $\alpha=-1.35\pm0.05$, therefore, galaxies
in groups have a steeper $\alpha$ in $\sim 0.12$. 
Nevertheless, owing to the errors in both determinations, the faint 
end slopes are statistically comparable ($\sim 1.5 \ \sigma$ difference).
It should be noted that the similarity between the results obtained for field and 
group galaxies in the Red sequence it is not unexpected since these 
galaxies are mainly located in galaxy systems. On the other hand, we do expect 
a difference for the LF of the Blue sequence in groups compared with
that obtained in the field, since a large fraction of Blue galaxies are not
in groups and, consequently, do not suffer the action of some typical physical processes
of dense environments. 

\begin{figure}
\plotone{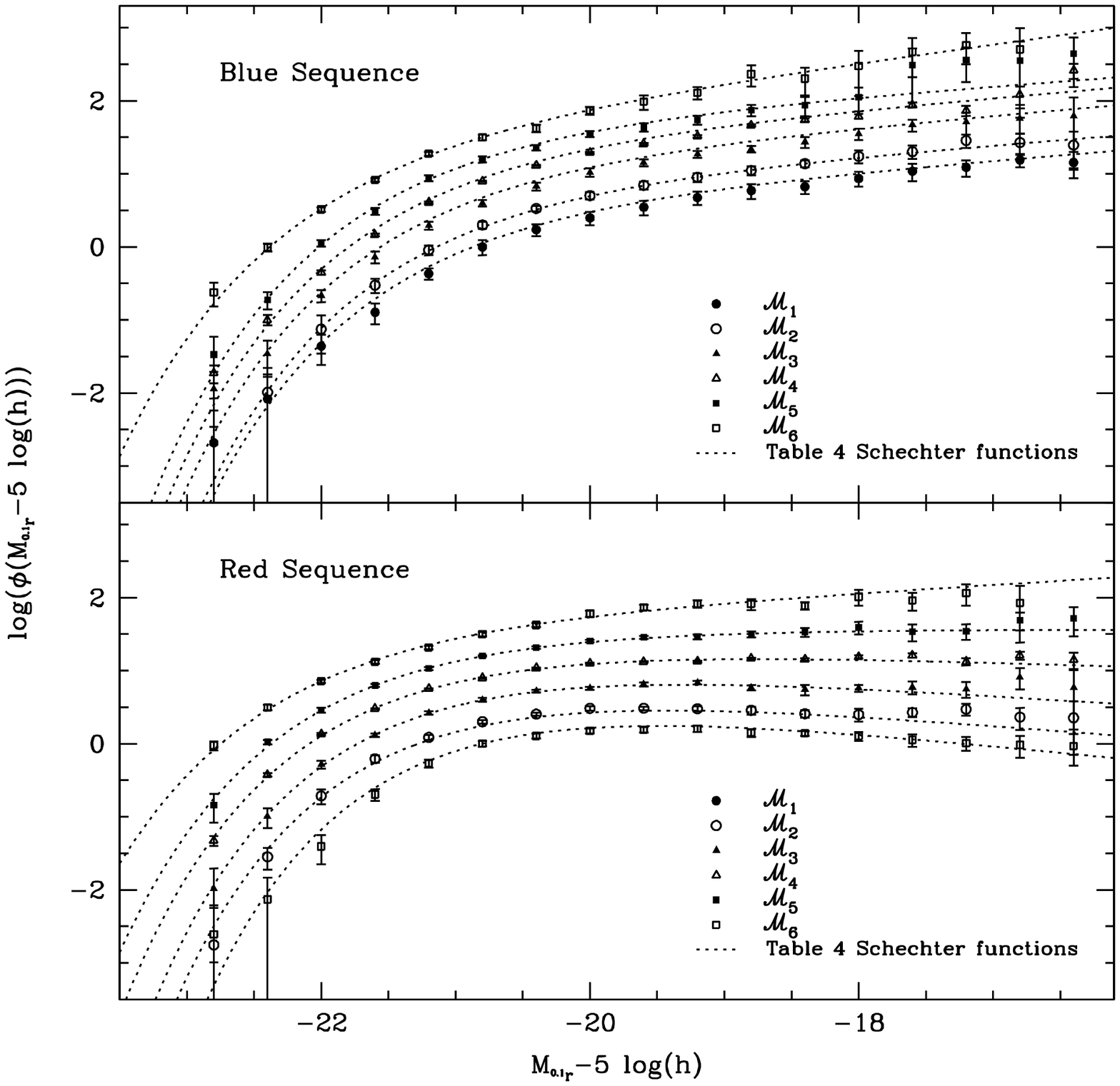}
\caption{$^{0.1}r$ band luminosity function of the Red (upper panel)
and Blue (lower panel) sequences in groups for different group mass ranges 
(arbitrary units). Dotted lines show
the best-fitting Schechter functions. 
Error bars are computed using a bootstrapping technique.
}
\label{lfmc}
\end{figure}

\begin{figure}
\plotone{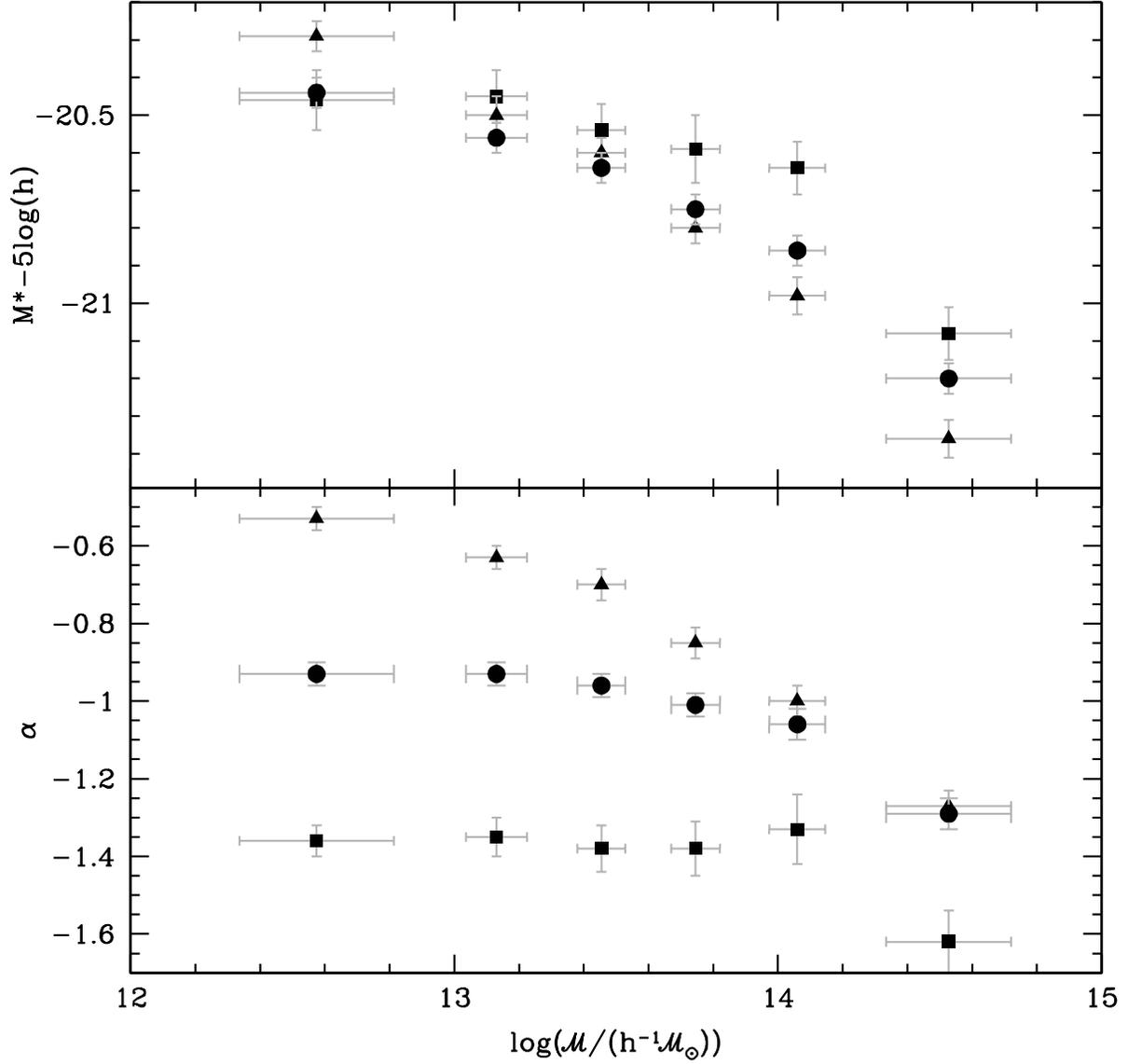}
\caption{
The best-fitting Schechter parameters for the $^{0.1}r$ 
band LF of red (triangles) and blue (squares) galaxies 
in groups as a function of the median mass of a given range. 
The mass error bars are the semi-interquartile range, whereas
the errors in the Schechter parameters are the projections 
of 1$\sigma$ joint error ellipse onto each axis.
Circles are the best-fitting Schechter parameters for the whole sample
of galaxies in groups in each mass range as plotted in 
Figure \ref{amm}.
}
\label{ammcolor}
\end{figure}

We have also computed the $\rb$ band LF for Blue and Red sequences as a function
of group mass, splitting groups into the same six mass bins used in the 
previous subsection. 
The resulting LFs are shown in Figure \ref{lfmc} while the corresponding best-fitting 
Schechter parameters are quoted in Table \ref{stymasscolor} and shown in 
Figure \ref{ammcolor}.
In the later, upper panel shows the behavior of $M^{\ast}$ as a function
of mass, meanwhile the lower panel displays $\alpha$ as a function of mass.
We also show in both panels the variation of the Schechter parameters  with mass
for all galaxies in groups, computed in the previous subsection.
Regarding the variation of $M^{\ast}$ with mass, it is clear that the change
is larger for the Red sequence. $M^{\ast}$ smoothly decreases in $\sim 1.1$ magnitudes 
between the lowest and the highest mass subsamples. For the Blue sequence,
$M^{\ast}$ shows a small variation ($\sim 0.2$ magnitudes) for the first 5 
mass bins, and doubles this brightening in the last one. 
The faint end slope as a function of mass also shows a larger variation
for the Red sequence, a steepening of 0.75 from low to high mass.  
Excluding the highest mass bin ($\alpha\sim-1.6$), 
the Blue sequence has a constant value of $\alpha\sim-1.4$.

According to our results, the Red sequence luminosities show a strong
variation with the system mass, while the Blue sequence luminosities 
are roughly independent of mass. In the previous subsection we found that
$M^{\ast}$ brightens and $\alpha$ steepens when system mass increases, hence
we can now conclude that those changes are mainly observed in one of the populations,
the Red one.

\subsection{The influence of the brightest galaxy}

When studying clusters of galaxies, one of the most interesting subjects 
is the Brightest Cluster Galaxy (BCG). They are preferentially elliptical, are located
in the center of the potential well and are particularly massive and bright.
Even though their evolution and the effect of the environment upon them are not
well understood, the most plausible scenario for their origin is that they form rapidly
from mergers of several galaxies during the early stages of cluster or group
collapse. Later, they become brighter in more massive systems as hierarchical
structure formation continues \citep{merritt,edge,dubinski,lin}. 

\begin{figure}
\plotone{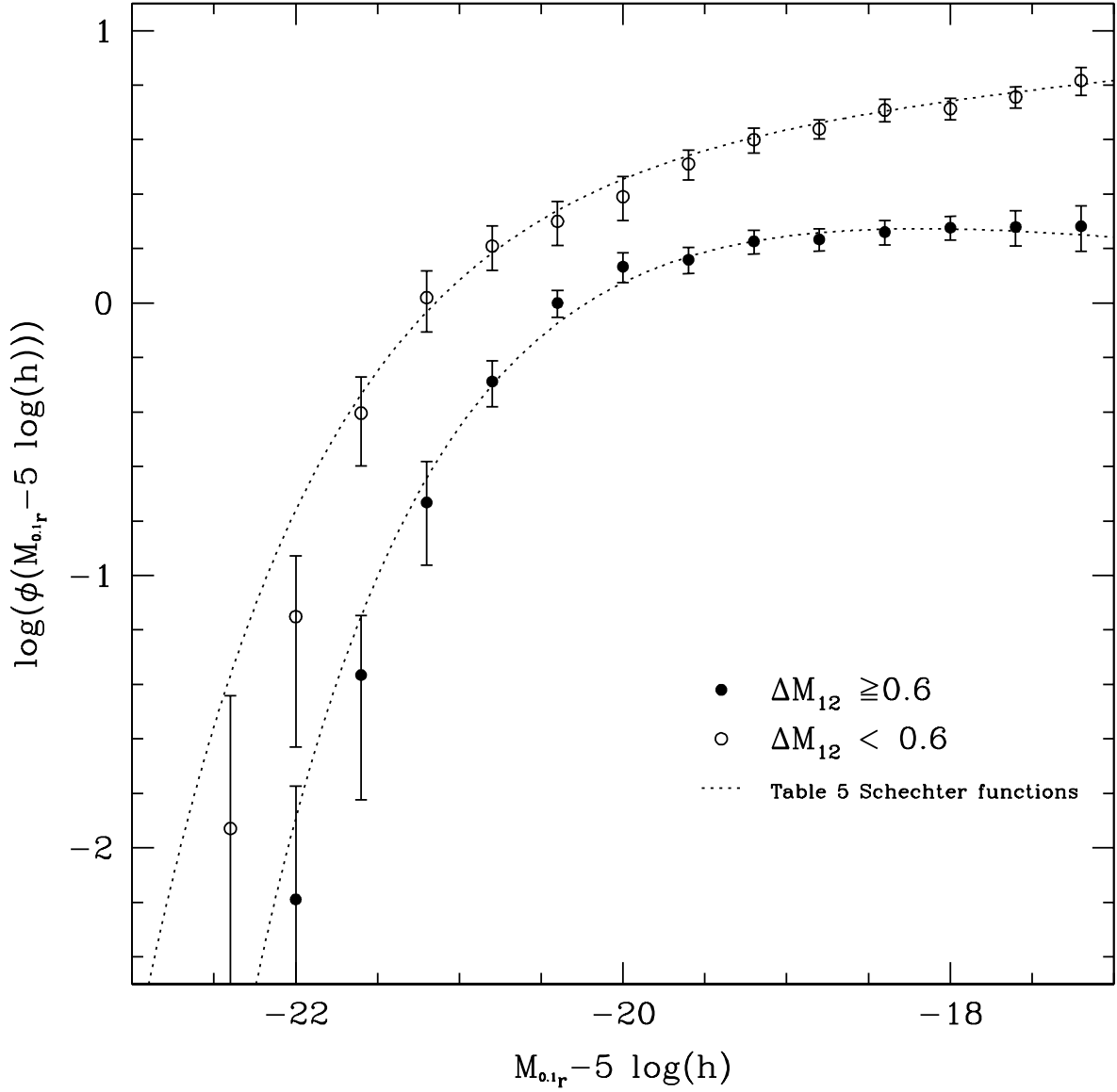}
\caption{
$^{0.1}r$ band luminosity functions of galaxies in groups for
samples with absolute magnitude difference between the 
first and second brightest galaxies $\Delta M_{12}$ greater or 
equal than 0.6 (upper panel) and lesser than 0.6 (lower panel) (arbitrary units).
Dotted lines show the best-fitting Schechter functions.
Error bars are computed using a bootstrapping technique.
}
\label{lfbgg}
\end{figure}

In this subsection we are interested in studying the LF of groups 
(i.e. low and intermediate mass systems) where there is an important luminosity 
difference between the brightest group galaxy (BGG) and the second ranked galaxy.
To do so, we have split the sample of groups into two subsamples with
roughly the same number of groups: 
those with a magnitude difference between the BGG and the second ranked galaxy,
$\Delta M_{12}\ge 0.6$, and those with $\Delta M_{12}<0.6$. 
It is worth to be mentioned that the absolute magnitude distribution of the BGGs
for both subsamples are quite similar.
In order to avoid any possible bias with redshift we have restricted the
group subsamples to $0.02\le z\le0.05$.
The resulting luminosity functions are shown in Figure \ref{lfbgg} and the
corresponding best-fitting Schechter parameters are in Table \ref{stybgg}.
For the $\Delta M_{12}\ge 0.6$ subsample $M^{\ast}$ is $\sim0.7$ magnitudes brighter,
and $\alpha$ is $\sim0.3$ steeper with $\sim 5 \sigma$ significance,  
than the $\Delta M_{12}< 0.6$ subsample values.
The two subsamples have similar group mass distributions. This was also found 
by \citet{lin} in clusters, where the difference in luminosity 
between the BCG and the second ranked galaxy do not correlate with mass. 
Hence, the results above, can not be associated with a possible mass bias. 

As said above, the most likely scenario for the 
BCGs formation is based upon the idea that these objects have been formed from 
mergers between several galaxies that take place in groups or low mass cluster.
Under this scheme, the merging 
galaxies that form the BCGs are the BGGs, i.e. the more luminous galaxies in groups. 
Since the merger rate is a decreasing function of the velocity dispersion
of a galaxy system, mergers should be the main responsible for the formation of the 
BGGs, progenitors of the future BCG. From our results, we conclude that 
BGGs that are considerably brighter than the remaining group members,
are preferentially found in groups where mergers have been more effective, producing
a brightening of the characteristic magnitude $M^{\ast}$.
Moreover, it is known that the galactic cannibalism solely at the present epoch 
is unlikely to be the process responsible for the BGG origin since this scenario
is not in agreement with a high value of $\Delta M_{12}$ \citep{merritt,tremaine}.
\citet{loh} have observed that the large values of $\Delta M_{12}$ measured
around the Luminous Red Galaxies in the SDSS support the idea 
that the BCGs form during the system collapse due to the process 
of merging small structures with a subsequent growth due to the accretion
of lesser members. Therefore, we can add to our picture that the BGGs of our subsample
with larger values of $\Delta M_{12}$ went through merging processes at the
early stages of the galaxy systems formation.

\section{Summary and discussion}
In this work we have studied the behavior of galaxy luminosities in groups of galaxies.
We use the largest sample of galaxy groups at the present which
allow us to obtain very reliable statistical results. The groups are identified in 
the SDSS DR4 using the same procedure as the described by \citet{mz05}. 
The galaxy luminosities analysis is performed computing the LF for several subsamples
of galaxies in groups. Our results can be summarized as follows:
\begin{itemize}
\item First, we compute the LFs of galaxies in groups in different SDSS bands. Our
results show that, except for the $\ub$ band, the characteristic magnitude, $M^{\ast}$, 
is $\sim0.5$ magnitudes brighter than the obtained for field galaxies by \citet{blantonlf}.
The faint end slope, $\alpha$, of these LFs are slightly steeper than their field 
counterparts. 
\item Then, we study the dependence of the $\rb$ LF of galaxies in groups for different
bins in group mass. We observe two clear trends: a brightening of the characteristic
magnitude ($\Delta M^{\ast} \sim 0.75$) and a steepening of the faint end slope 
($\Delta\alpha \sim 0.4$) as mass increases. 
\item Similar results are found when analyzing the mass dependence of the LFs
in the remaining SDSS bands for both, $M^{\ast}$ and $\alpha$. Only the $\ub$ band
shows a less pronounced brightening in the characteristic magnitude. 
Since the $\ub$ band is more sensitive to current galaxy star formation, we expect 
that the LFs in this band to be less suitable to trace the mass, i.e., to show the real 
brightening of $M^{\ast}$ as a function of group mass as observed in the other bands. 
\item We made use of the bimodal $^{0.1}(u-r)$ distribution to split the galaxy sample
into a Red and a Blue sequence. The divider point among the populations is estimated
from a linear relation as a function of the $\rb$ absolute magnitude. The LF obtained
for the Red sequence is quite similar to that obtained by \citet{baldry03} for 
field galaxies. On the other hand, even when the Blue sequence in groups shows 
a similar faint end slope than the observed for the Blue sequence in the field, 
its characteristic magnitude is significantly brighter than the field counterpart.   
\item Regarding the dependence of Red and Blue sequence LF in groups with the group
mass we observe that the Red sequence shows stronger changes in both, the 
characteristic magnitude and the faint end slope. The behavior of the Blue sequence
remains almost unchanged as system mass increases.
\item Finally, studying the effect of the brightest group galaxy on these environments
we split the galaxy sample into two subsamples: groups with ($\Delta M_{12}\ge 0.6$) and 
without ($\Delta M_{12} < 0.6$) a remarkable luminosity difference between the brightest 
group galaxy and the second ranked galaxy. We observe that those groups with larger 
$\Delta M_{12}$ have a brighter $M^{\ast}$ and a steeper $\alpha$ than
the other group subsample.
\end{itemize} 

The most plausible scenario to explain the results above is one in which mergers 
have played a fundamental role in galaxy evolution in groups. 
It is well known from the literature that galaxy mergers are more frequent in low 
mass systems since the low galaxy velocity dispersion should induce higher 
encounter rates \citep{merritt}. 
They can account for the observed brightening of $M^{\ast}$ with mass and the reddening 
of galaxy colors given that these processes could consume gas in a burst of star formation 
and subsequently, at later times, induce lower star formation rates. 
Our results fit into the scenario proposed by \citet{baldry03} where mergers are the 
cause of the color bimodality, with a red population resulting from merger processes 
and a blue population that form stars at a rate determined by their internal physical
properties. 
The fact that the LF of the blue population in groups remains almost unchanged
with mass, supports the idea of a population that evolves independently of environment,
while the observed behavior in the LF of the red galaxies reveals that these objects have 
been through major changes due to environmental effects. 
Recently, \citet{faber} have studied the evolution of the LF of the
Red and Blue sequences up to $z\sim1$ using also the bimodality distribution
to perform the distinction among both populations. The main result is that the
number density obtained for Red sequence galaxies shows a strong evolution, while the Blue
sequence remains almost unchanged in the sampled redshift range. They conclude
that the more plausible scenario to account for this effect is one in which
some blue galaxies have suffer a strong suppression of their star formation due
to gas-rich mergers, resulting in a migration to the red sequence 
and subsequently evolving there due to several stellar mergers.
The \citet{faber} model is quite consistent with the one suggested here to explain
the behaviour of both populations. 

Another result that should be explained is the decrease of the faint end slope with
group mass. This parameter is the one that describes the shape of the Schechter function
and it is defined by both, the faint and the bright ends of the LF. 
Therefore, any significant variation in any of these regions, should change the value of 
$\alpha$. A shallower distribution of bright galaxies can result in a steeper value
of $\alpha$ without changing the number of faint galaxies. Hence, the merger scenery 
can also explain the resulting behavior of $\alpha$ with mass.
This does not exclude the possible incidence of environmental processes such as galaxy
harassment, ram pressure, etc., that could enhance the number of faint galaxies.
Nevertheless, it should be reminded that the limiting apparent magnitude of the MGS 
does not allow to probe the very faint end of the LF, therefore, the observed change in 
$\alpha$ with mass, can not be unambiguously associated with a dwarf galaxy population.

\section*{Acknowledgments}
This work has been partially supported by grants from 
Consejo de Investigaciones 
Cient\'{\i}ficas y T\'ecnicas de la Rep\'ublica Argentina (CONICET), the
Secretar\'{\i}a de Ciencia y T\'ecnica de la Universidad Nacional de C\'ordoba
(SeCyT), Agencia Nacional de Promoci\'on Cient\'\i fica de la Rep\'ublica
Argentina and Agencia C\'ordoba Ciencia.

Funding for the Sloan Digital Sky Survey (SDSS) has been provided by the 
Alfred P. Sloan 
Foundation, the Participating Institutions, the National Aeronautics and Space 
Administration, the National Science Foundation, the U.S. Department of Energy, 
the Japanese Monbukagakusho, and the Max Planck Society. The SDSS Web site is 
http://www.sdss.org/.
The SDSS is managed by the Astrophysical Research Consortium (ARC) for the 
Participating Institutions. The Participating Institutions are The University 
of Chicago, Fermilab, the Institute for Advanced Study, the Japan Participation 
Group, The Johns Hopkins University, the Korean Scientist Group, Los Alamos 
National Laboratory, the Max Planck Institut f\"ur Astronomie (MPIA), the 
Max Planck Institut f\"ur Astrophysik (MPA), New Mexico State University, 
University of Pittsburgh, University of Portsmouth, Princeton University, 
the United States Naval Observatory, and the University of Washington.

\newpage

\begin{deluxetable}{lccccc}
\tablecaption{STY best-fitting Schechter parameters for the luminosity
functions of galaxies in groups \label{sty}}
\tablehead{
\colhead{Band} & \colhead{${\rm m_{lim}}$} & \colhead{Redshift range} &
\colhead{$N_{\rm galaxies}$} & \colhead{$M^{\ast}-5\log(h)$} & 
\colhead{$\alpha$}
}
\startdata
$^{0.1}u$ & 18.40 & $0.02-0.14$ & 15743 & $-17.83\pm0.03$ & $-1.13\pm0.04$\\
$^{0.1}g$ & 17.70 & $0.02-0.17$ & 37239 & $-19.71\pm0.02$ & $-1.09\pm0.02$\\
$^{0.1}r$ & 17.77 & $0.02-0.22$ & 83869 & $-20.85\pm0.01$ & $-1.08\pm0.01$\\
$^{0.1}i$ & 16.91 & $0.02-0.22$ & 54233 & $-21.21\pm0.02$ & $-1.15\pm0.02$\\
$^{0.1}z$ & 16.50 & $0.02-0.22$ & 46001 & $-21.47\pm0.02$ & $-1.12\pm0.01$\\
\enddata
\end{deluxetable}

\begin{deluxetable}{lccccc}
\tablecaption{Group mass dependence of the LF in the $\rb$ band: STY best-fitting Schechter parameters 
\label{stymass}}
\tablehead{
\colhead{Mass sample} & \colhead{Mass range\tablenotemark{a}} & 
\colhead{$N_{\rm groups}$} & 
\colhead{$N_{\rm galaxies}$} & \colhead{$M^{\ast}-5\log(h)$} & 
\colhead{$\alpha$} 
}
\startdata
${\cal M}_1$ & $11.0-12.9$ & 2322 & 10326 & $-20.44\pm0.04$ & $-0.93\pm0.03$\\
${\cal M}_2$ & $12.9-13.3$ & 2392 & 12514 & $-20.56\pm0.04$ & $-0.93\pm0.03$\\
${\cal M}_3$ & $13.3-13.6$ & 2409 & 13580 & $-20.64\pm0.04$ & $-0.96\pm0.03$\\
${\cal M}_4$ & $13.6-13.9$ & 2362 & 14415 & $-20.75\pm0.04$ & $-1.01\pm0.03$\\
${\cal M}_5$ & $13.9-14.25$ & 2150& 14626 & $-20.86\pm0.04$ & $-1.06\pm0.04$\\
${\cal M}_6$ & $>14.25$ & 2231 & 17331 & $-21.20\pm0.04$ & $-1.29\pm0.04$\\
\enddata
\tablenotetext{a}{Units are $\log({\cal M}/(h^{-1}M_{\odot}))$}
\end{deluxetable}

\begin{deluxetable}{lccc}
\tablecaption{Group mass dependence of the LF in the
$\ub$, $\gb$, $\ib$, and $\zb$ bands: STY best-fitting Schechter parameters\label{stymassband}}
\tablehead{
\colhead{Mass sample\tablenotemark{a}} & 
\colhead{$N_{\rm galaxies}$} & 
\colhead{$M^{\ast}-5\log(h)$} & 
\colhead{$\alpha$} 
}
\startdata
\cutinhead{$\ub$ band}
${\cal M}_1$ & 4067 & $-17.72\pm0.06$ & $-1.03\pm0.07$\\
${\cal M}_2$ & 3727 & $-17.70\pm0.06$ & $-0.98\pm0.08$\\
${\cal M}_3$ & 3025 & $-17.77\pm0.06$ & $-1.07\pm0.10$\\
${\cal M}_4$ & 2220 & $-17.91\pm0.09$ & $-1.15\pm0.12$\\
${\cal M}_5$ & 1460 & $-18.10\pm0.09$ & $-1.51\pm0.12$\\
${\cal M}_6$ & 1008 & $-18.15\pm0.09$ & $-1.55\pm0.14$\\
\cutinhead{$\gb$ band}
${\cal M}_1$ & 6794 & $-19.42\pm0.04$ & $-0.93\pm0.04$\\
${\cal M}_2$ & 7303 & $-19.49\pm0.04$ & $-0.91\pm0.05$\\
${\cal M}_3$ & 6965 & $-19.57\pm0.05$ & $-0.95\pm0.05$\\
${\cal M}_4$ & 6381 & $-19.77\pm0.05$ & $-1.13\pm0.06$\\
${\cal M}_5$ & 5209 & $-19.90\pm0.06$ & $-1.27\pm0.06$\\
${\cal M}_6$ & 4095 & $-20.12\pm0.07$ & $-1.41\pm0.07$\\
\cutinhead{$\ib$ band}
${\cal M}_1$ & 7559 & $-20.74\pm0.05$ & $-0.90\pm0.04$\\
${\cal M}_2$ & 8842 & $-20.93\pm0.05$ & $-0.95\pm0.04$\\
${\cal M}_3$ & 9266 & $-21.02\pm0.04$ & $-1.02\pm0.04$\\
${\cal M}_4$ & 9457 & $-21.16\pm0.05$ & $-1.12\pm0.05$\\
${\cal M}_5$ & 9060 & $-21.30\pm0.05$ & $-1.26\pm0.05$\\
${\cal M}_6$ & 9528 & $-21.64\pm0.05$ & $-1.50\pm0.05$\\
\cutinhead{$\zb$ band}
${\cal M}_1$ & 6519  & $-21.01\pm0.05$ & $-0.87\pm0.05$\\
${\cal M}_2$ & 7643 & $-21.20\pm0.05$ & $-0.91\pm0.05$\\
${\cal M}_3$ & 7938 & $-21.28\pm0.05$ & $-0.98\pm0.05$\\
${\cal M}_4$ & 8092 & $-21.43\pm0.05$ & $-1.09\pm0.05$\\
${\cal M}_5$ & 7619 & $-21.57\pm0.05$ & $-1.23\pm0.06$\\
${\cal M}_6$ & 7861 & $-21.87\pm0.05$ & $-1.45\pm0.05$\\
\enddata
\tablenotetext{a}{See Table \ref{stymass}}
\end{deluxetable}

\begin{deluxetable}{lccc}
\tablecaption{Group mass dependence of the LF in the $\rb$ band for Blue and Red sequences
\label{stymasscolor}}
\tablehead{
\colhead{Mass sample\tablenotemark{a}} & 
\colhead{$N_{\rm galaxies}$} & 
\colhead{$M^{\ast}-5\log(h)$} & 
\colhead{$\alpha$} 
}
\startdata
\cutinhead{Blue sequence}
All masses   &  29896 & $-20.76\pm0.03$ & $-1.47\pm0.03$\\
${\cal M}_1$ &   4542 & $-20.46\pm0.08$ & $-1.36\pm0.04$\\
${\cal M}_2$ &   4820 & $-20.45\pm0.07$ & $-1.35\pm0.05$\\
${\cal M}_3$ &   4922 & $-20.54\pm0.07$ & $-1.38\pm0.06$\\
${\cal M}_4$ &   4804 & $-20.59\pm0.09$ & $-1.38\pm0.07$\\
${\cal M}_5$ &   4514 & $-20.64\pm0.07$ & $-1.33\pm0.09$\\
${\cal M}_6$ &   5785 & $-21.08\pm0.07$ & $-1.62\pm0.08$\\
\cutinhead{Red sequence}
All masses   & 53941 & $-20.85\pm0.02$ & $-0.87\pm0.02$\\
${\cal M}_1$ &  5776 & $-20.29\pm0.04$ & $-0.53\pm0.03$\\
${\cal M}_2$ &  7680 & $-20.50\pm0.05$ & $-0.63\pm0.03$\\
${\cal M}_3$ &  8655 & $-20.60\pm0.04$ & $-0.70\pm0.04$\\
${\cal M}_4$ &  9608 & $-20.80\pm0.04$ & $-0.85\pm0.04$\\
${\cal M}_5$ & 10110 & $-20.98\pm0.05$ & $-1.00\pm0.04$\\
${\cal M}_6$ & 11544 & $-21.36\pm0.05$ & $-1.27\pm0.04$\\
\enddata
\tablenotetext{a}{See Table \ref{stymass}}
\end{deluxetable}

\begin{deluxetable}{lcccc}
\tablecaption{LF of galaxies in groups splitted by the difference in magnitude
between the brightest and the second brightest galaxy\label{stybgg}}
\tablehead{
\colhead{$\Delta M_{12}$} & 
\colhead{Redshift range} & 
\colhead{$N_{\rm galaxies}$} & 
\colhead{$M^{\ast}-5\log(h)$} & 
\colhead{$\alpha$}
}
\startdata
$\ge 0.6$ & $0.02-0.05$ & 6447 & $-20.8\pm0.1$ & $-1.14\pm0.03$\\
$< 0.6$   & $0.02-0.05$ & 6553 & $-20.12\pm0.06$ & $-0.83\pm0.03$\\
\enddata
\end{deluxetable}

\end{document}